\documentclass[doublecol]{epl2}
\usepackage{graphics,graphicx,epsfig,color,hyperref,amsmath,amssymb,latexsym,float}
\usepackage{times}
\input pdfcolor.tex

\title{Mott transition and anomalous resistive state \\ in the pyrochlore molybdates}
\shorttitle{Mott transition, pyrochlore molybdates}

\author{Nyayabanta Swain \and Pinaki Majumdar}
\shortauthor{Swain and Majumdar}

\institute{ 
Harish-Chandra Research Institute, HBNI, 
Chhatnag Road, Jhunsi, Allahabad 211019, India
}


\abstract{
The rare-earth based pyrochlore molybdates involve orbitally degenerate 
electrons Hund's coupled to local moments. The large Hund's coupling 
promotes ferromagnetism, the superexchange between the local moments 
prefers antiferromagnetism, and Hubbard repulsion  tries to open a Mott 
gap. The phase competition is tuned by the rare-earth ionic radius, 
decreasing which leads to change from a ferromagnetic metal to a spin disordered 
highly resistive ground state, and ultimately an `Anderson-Mott' insulator. 
We attempt a quantitative theory  of the molybdates by studying their minimal 
model on a pyrochlore geometry, using a static auxiliary field based
Monte Carlo. We establish a thermal phase diagram 
that closely corresponds to the experiments, predict the hitherto unexplored 
orbital correlations, quantify and explain the origin of the anomalous 
resistivity, and present dynamical properties across the metal-insulator 
transition.  
}

\begin{document}

\maketitle

\section{Introduction}

Traditional Mott materials involve a strong on-site Coulomb
interaction that, beyond a critical value, and at integer 
filling, inhibits electron motion \cite{MIT_Mott}. 
This, in a clean material,
leads to an abrupt change in the zero temperature state 
from perfectly conducting to non conducting. The non
conducting state typically has strong antiferromagnetic
(AF) correlations, if not long range order, 
since that lowers the kinetic energy.

The Mott transition on a frustrated structure brings in a 
novelty since the AF ordered state in the Mott phase 
cannot be realised and one may have the signatures of a 
`spin liquid' \cite{SL-triangle,SL-fr-mag}. 
Such phases are realised in some triangular 
lattice organics \cite{SL-tr-organic1,SL-tr-organic2,SL-tr-organic}. 
The pyrochlores \cite{pyr-rmp} are also highly
frustrated structures, much studied for possible spin 
liquid phases, but the rare earth molybdates, 
R$_2$Mo$_2$O$_7$, add  additional twists
to the Mott problem: 
(i)~the Mott transition in these materials occur 
in the background of overall {\it ferromagnetic} correlation 
\cite{Mo-var-mag-ele1,Mo-var-mag-ele2,pyr-rmp}, 
and (ii)~the zero temperature resistivity
seems to grow continuously with the control parameter
\cite{Mo-optics2} (see next) 
rather than having an abrupt zero to infinity transition.
These features owe their origin to the additional 
degrees of freedom, and couplings, involved in these
materials.

The R$_2$Mo$_2$O$_7$ family exhibit ground states that vary 
from a ferromagnetic metal (FM) to a spin glass metal (SG-M) 
and then a spin glass insulator (SG-I) 
as the rare earth radius $r_R$ is reduced
\cite{Mo-MIT-expt2}.
Materials with R = Nd and Sm are metallic,  
R = Tb, Dy, Ho, Er, and Y are insulating, 
and R=Gd is on the verge of the insulator-metal transition (IMT)
\cite{Mo-var-mag-ele2,Mo-optics2,Mo-optics4}.
The highest observed ferromagnetic $T_c$ is $\sim 100$K, in Nd, 
while the spin glass transition temperature, $T_{SG}$ 
is typically $\sim 20$K \cite{Mo-Tc-Tf1,Mo-Tc-Tf2,Mo-Tc-Tf3}. 
The unusual features in transport include
very large residual resistivity, $\sim$ 10 $m\Omega$cm
close to the metal-insulator transition \cite{Mo-optics2},
prominent anomalous Hall effect in metallic samples,
{\it e.g}, Nd$_2$Mo$_2$O$_7$
\cite{Mo-anm-hall1,Mo-anm-hall2,Mo-anm-hall3,Mo-anm-hall4,Mo-anm-hall5},
and magnetic field
driven metallisation in the weakly insulating samples, 
{\it e.g}, Gd$_2$Mo$_2$O$_7$ \cite{hanasaki-andmott}.

The qualitative physics behind these effects is not a mystery, but 
several major questions remain unanswered, {\it e.g}, on 
(i)~the simultaneity of the metal-insulator and magnetic transition
in the ground state, (ii)~the thermal scales for the 
magnetic transitions, (iii)~the orbital state, whose fate is 
entangled with the spin state, and 
(iv)~the transport near the Mott transition.

We employ a real space approach, using a static auxiliary orbital
field to handle the Hubbard interaction, and solve the
resulting `electron~-~local moment~-~orbital moment' problem 
via a Monte Carlo scheme on the pyrochlore lattice. After making
parameter choices suggested by {\it ab initio} estimates 
\cite{molybdate-DFT1,molybdate-DFT2}
our main results are as follows.

(i)~We obtain a phase diagram with ferromagnetic metal and
spin disordered metal and insulator phases. However, our 
disordered state is a `spin liquid' in contrast
to the experimental spin glass.
(ii)~The FM $T_c$ is in the experimental ballpark if we
make room for a simple renormalisation of the Hund's coupling.
(iii)~Our transport results bear almost quantitative 
correspondence with experiments \cite{Mo-optics2,Mo-optics4}, 
explain the high residual resistivity
in terms of spin and orbital disorder scattering, and
predict a highly non monotonic temperature dependence
for samples of the form Gd$_{2-x}$Sm$_x$Mo$_2$O$_7$.
(iv)~The temperature and correlation dependence of our 
intermediate frequency optical spectral weight is very similar
to the experiments \cite{Mo-optics2,Mo-optics4}
but changing the cutoff frequency
reveals peculiarities, also reflected in the
single-particle weight.

\section{Model and method}

The R$_2$Mo$_2$O$_7$ 
structure consists of two interpenetrating 
pyrochlore lattices,
one formed by Mo cations and the other by R.  
The Mo atom has octahedral oxygen coordination. The 
resulting crystal field splits the fivefold  
degenerate Mo 4d states into doubly degenerate $e_g$
and  triply degenerate $t_{2g}$ manifolds,
and a trigonal distortion
splits the $t_{2g}$ further into a nondegenerate
$a_{1g}$ and a doubly degenerate $e'_g$ \cite{molybdate-DFT1}. 
The Mo cation is 
nominally tetravalent and has two electrons on average.
The deeper  $a_{1g}$ electron behaves like a local moment, 
and the single electron in the two  $e'_g$ orbitals 
is the `itinerant' degree of freedom \cite{molybdate-DFT1}.  
The $e_g$ states remain unoccupied.
The accepted model \cite{Molyb-init-model}
for these degrees of freedom takes the following form:
\begin{eqnarray}
H & = &\sum_{\langle ij \rangle,\sigma}^{\alpha\beta}
t_{ij}^{\alpha\beta} 
c^{\dagger}_{i\alpha\sigma}c_{j\beta\sigma}  
- J_H \sum_{i,\alpha} {\bf S}_i.c^{\dagger}_{i\alpha\sigma}  
\vec{\sigma}_{\sigma\sigma^{'}} c_{i\alpha\sigma^{'}} \cr
&& 
~~ + H_{AF} + \sum_{i,\alpha\beta\alpha^{'}\beta^{'}}^{\sigma, \sigma'}
 U_{\alpha\beta}^{\alpha^{'}\beta^{'}}
 	c^{\dagger}_{i\alpha\sigma}c^{\dagger}_{i\beta\sigma^{'}}
c_{i\beta^{'}\sigma^{'}}c_{i\alpha^{'}\sigma} \nonumber  
\end{eqnarray}
where $H_{AF} = 
 J_{AF} \sum_{\langle ij \rangle} {\bf S}_i.{\bf S}_{j} $.
The first term is the kinetic energy, involving nearest neighbour 
intra and inter-orbital $e'_g$ hopping.
The second term is the Hund's coupling 
between the $a_{1g}$ local moment ${\bf S}_i$ and the $e'_g$ 
electrons,
$J_{AF}$ is the AF superexchange coupling between local moments at
neighbouring sites on the pyrochlore lattice, and the $U$ 
represents onsite $e'_g$ Coulomb matrix elements.

To simplify the computational problem 
we treat the localized spins ${\bf S}_i$ 
as classical unit vectors, absorbing the size $S$ in
the magnetic couplings.
Also, to reduce the size of the Hilbert space
we assume that $J_H/t \gg 1$, where $t$ is the
typical hopping scale, so that only the
locally `spin aligned' fermion state is retained.
In this local basis the hopping matrix elements 
are dictated by the orientation of the ${\bf S}_i$ 
on neighbouring sites. 
These lead to a simpler model:
$$
H  =  \sum_{\langle ij \rangle,\alpha\beta}
{\tilde t}_{ij}^{\alpha\beta} {\tilde c}^{\dagger}_{i\alpha} {\tilde c}_{j\beta} 
   + J_{AF} \sum_{\langle ij \rangle} {\bf S}_i.{\bf S}_j
+ U \sum_{i}^{\alpha \neq \beta} n_{i\alpha} n_{i\beta} 
$$
where the fermions are now `spinless'.
$U >0$ is the inter-orbital Hubbard repulsion. 
The effective hopping is determined by the orientation of
the localized spins. If  ${\bf S}_i = (sin\theta_{i}
cos\phi_{i},sin\theta_{i}sin\phi_{i},cos\theta_{i})$ and
${\bf S}_j = (sin\theta_{j}
cos\phi_{j},sin\theta_{j}sin\phi_{j},cos\theta_{j})$
then $t_{ij}^{\alpha\beta} = [ cos\frac{\theta_i}{2}
cos\frac{\theta_j}{2}
		 + sin\frac{\theta_i}{2}sin\frac{\theta_j}{2} 
e^{-i(\phi_i - \phi_j)}] t^{\alpha\beta} $,
with $t^{11} = t^{22} =t$ and $t^{12} = t^{21} =t'$.
We set $ t' = 1.5t$ as suggested by the density functional theory
\cite{molybdate-DFT2} and keep only nearest neighbour hopping.

The first two terms represent fermions in a classical spin
background and the resulting magnetic phase competition has
been studied on a pyrochlore lattice \cite{Molyb-DE-SE}. 
While these results are interesting 
they miss out on the large correlation scale,
$U$, that drives the Mott transition. One option
is to treat the model within dynamical mean field theory 
(DMFT) \cite{DMFT-georges}, 
but then the spatial character crucial to the pyrochlore 
lattice is lost. 

We opt to handle the problem in real space
as follows: (i)~We use a Hubbard-Stratonovich (HS)
\cite{hubb-strat,hubbard,schulz}
transformation 
that decouples $U n_{i \alpha} n_{i \beta}$ in terms of
an auxiliary orbital moment ${\bf \Gamma}_i(\tau)$,
coupling to ${\bf O}_i = \sum_{\mu \nu} c^{\dagger}_{i \mu}
{\vec \sigma}_{\mu  \nu} c_{i \nu}$, and 
a scalar field $\Phi_i(\tau)$ coupling to $n_i$ at each site
\cite{method_SPA_detail}.
(ii)~An exact treatment of the resulting functional integral 
requires quantum Monte Carlo. Here we retain 
only the zero Matsubara frequency modes of 
${\bf \Gamma}_i$ and $\Phi_i$, {\it i.e}, approximate
them as classical fields. (iii)~The spatial thermal
fluctuations of ${\bf \Gamma}_i$ are completely retained, 
while $\Phi_i$ is treated at the saddle point level, 
setting $\Phi_i \rightarrow \langle \Phi_i \rangle = (U/2)
\langle n_i \rangle = U/2$ at half-filling
(since charge fluctuations are expensive at large $U$).
This method overall is known as a `static path approximation'
(SPA) to the functional integral for the many body partition
function and has been used earlier in several problems
\cite{SPA_dagotto,SPA_meir,mott-Tr}.
This leads to a more tractable problem.
\begin{eqnarray}
H_{eff}\{{\bf S}_i, {\bf \Gamma}_i\} &=& 
-\frac{1}{\beta} logTr e^{-\beta H_{el}} 
+ H_{AF} 
+ \frac{U}{4} \sum_{i} {\bf \Gamma}_{i}^2
\cr
H_{el}\{{\bf S}_i, {\bf \Gamma}_i\} & = & 
\sum_{ij}^{\alpha\beta} {\tilde t}_{ij}^{\alpha\beta} 
{\tilde c}^{\dagger}_{i\alpha}{\tilde c}_{j\beta}  - {\tilde \mu} \sum_i n_i 
- \frac{U}{2}\sum_{i}{\bf \Gamma}_{i}.{\bf O}_i  
\nonumber
\end{eqnarray}
with ${\tilde \mu} = \mu - U/2$, $\mu$ being the chemical potential.
  The localized spin and orbital moment configurations  
follow the distribution
$$
P\{{\bf S}_i, {\bf \Gamma}_i\} \propto 
\textrm{Tr}_{cc^{\dagger}} e^{-\beta H_{eff} }
$$
Within the SPA scheme $H_{eff}\{{\bf S}_i, {\bf \Gamma}_i\}$ 
and $P\{{\bf S}_i,{\bf \Gamma}_i\}$ 
define a coupled `fermion~-~local moment~-~orbital moment' problem. 

There are regimes where some analytic progress can be made,
but our results here are based on a Monte Carlo
solution of the model above $-$ generating the equilibrium
configurations of $\{ {\bf S}_i, {\bf \Gamma}_i\}$
through iterative diagonalisation of $H_{eff}\{{\bf S}_i, {\bf \Gamma}_i\}$.
To access large sizes within reasonable time we
use a cluster algorithm for estimating the update cost 
\cite{tca-sanjeev-pinaki,tca-anamitra-dagotto}.
Results in this paper are for a $6 \times 6 \times 6$ 
pyrochlore lattice (864 sites), using a cluster of
$3 \times 3 \times 3$ pyrochlore unit cells (108 sites).

From the equilibrium configurations we 
calculate the thermally averaged magnetic structure factor 
$S_{mag}({\bf q}) = \frac{1}{N^2}\sum_{ij}\langle{\bf S}_i.{\bf S}_j
\rangle e^{i{\bf q}.({\bf r}_i-{\bf r}_j)}$
and orbital structure factor 
$S_{orb}({\bf q}) = 
\frac{1}{N^2}\sum_{ij}\langle{\bf \Gamma}_i.{\bf \Gamma}_j
\rangle e^{i{\bf q}.({\bf r}_i-{\bf r}_j)}$
at each temperature. 

Electronic properties are calculated by diagonalising 
$H_{eff}\{{\bf S}_i, {\bf \Gamma}_i\}$ 
on the $6 \times 6 \times 6$ lattice on equilibrium backgrounds.
The optical conductivity for the molybdates is
calculated by using the Kubo formula \cite{kubo_allen} as follows,
\begin{eqnarray}
\sigma^{xx}(\omega)=\frac{\sigma_{0}}{N}
\langle \sum_{n,m}
{ {f(\epsilon_{n})- f(\epsilon_{m})}
\over {\epsilon_{m}-\epsilon_{n}} }
|J^{nm}_x|^2 \delta(\omega- (\epsilon_{m}-\epsilon_{n}))
\rangle
\nonumber
\end{eqnarray}
where $J^{nm}_x = \langle n \vert J_x \vert m \rangle$
and the current operator $J_x$ is given by,
\begin{eqnarray}
J_x&=&-i\sum_{i,\alpha\beta}\left[({\tilde t}_{i,i+\hat{x}}^{\alpha\beta}
{\tilde c}^{\dagger}_{i,\alpha}{\tilde c}_{i+\hat{x},\beta}-\textrm{h.c.})\right]
\nonumber
\end{eqnarray}
$f(\epsilon_{n})$ is the Fermi function,
$\epsilon_{n}$ and $|n\rangle$ are 
the single particle eigenvalues and eigenstates of
$H_{el}\{{\bf S}_i,{\bf \Gamma}_i\}$ respectively.
The conductivity is in units of
$\sigma_{0} = {e^2}/(\hbar a_0)$, 
where $a_0$ is the lattice constant. 
N is the total number of lattice sites.
The d.c. conductivity is obtained as a low frequency average of the
optical conductivity over a window $\sim 0.05t$.

We study the Mott transition in the molybdate family
with changing ionic radius of the rare-earth cation ($r_R$)  
by tuning the $U/t$ in our model.

\section{Results}

We will discuss the physics of the model for a wide
range of $J_{AF}$-$U$-$T$ in a separate paper and focus here
on parameters appropriate to the molybdates.
Following {\it ab initio} estimates 
\cite{molybdate-DFT1,molybdate-DFT2}
we use $t=0.1$ eV and $J_{AF} = 0.02$eV.
The calibration of  $U/t$ in terms of $r_R$ is 
based on the optical gap. We discuss this here briefly.

\begin{figure}[ht!]
\centerline{
\includegraphics[angle=0,width=7.5cm,height=8cm]{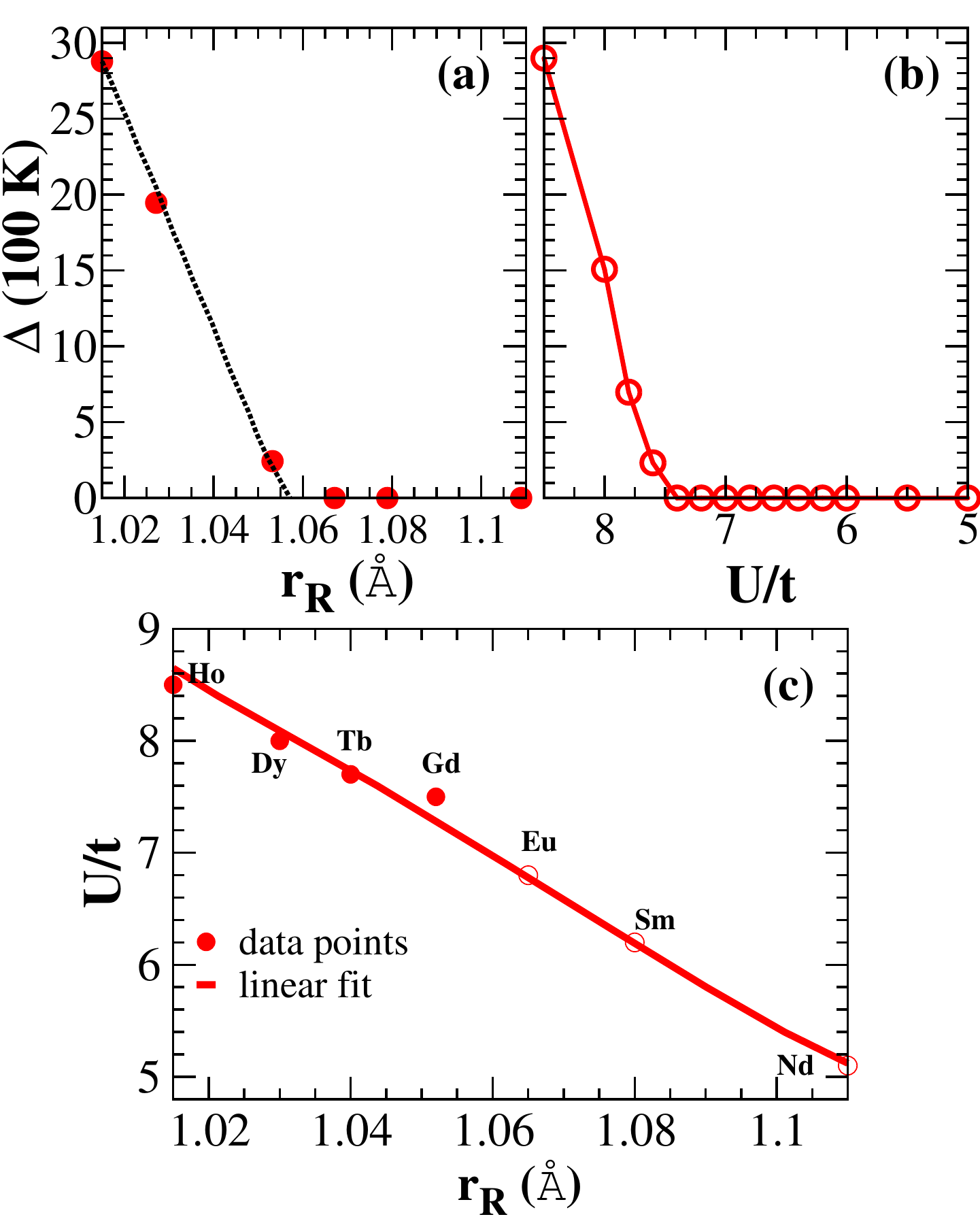}
}
\caption{\label{gap}Optical gap $\Delta$ extracted from the
low temperature optical conductivity:
(a)~experiment \cite{Mo-optics4} $\Delta_{expt}=\Delta (r_R)$
and (b)~theory $\Delta_{th}=\Delta (U/t)$.
(c)~By comparing the experimental optical gap, $\Delta(r_R)$,
with that of the theoretical optical gap, $\Delta(U/t)$, we estimate
the $U/t$ values appropriate for the rare-earth molybdates.
}
\end{figure}

{\it Parameter calibration:} 
In our calculation, the pyrochlore lattice constant $a_0 = a/4$, 
where $a$ is the FCC lattice constant.
For molybdates we have $a \sim 1.0$ nm \cite{molybdate-DFT1} 
and thus $\sigma_{0} \sim 10^4 (\Omega cm)^{-1}$. 
For $T \sim 0$, the optical gap $\Delta$ is determined 
by linearly extrapolating the decreasing edge of the 
optical conductivity spectra in the low energy regime.
Fig.\ref{gap}.(a) shows the comparison of experimental gap 
\cite{Mo-optics4} $\Delta_{expt}=\Delta (r_R)$ 
and theoretical gap $\Delta_{th}=\Delta (U/t)$ 
(see figure \ref{gap}.(b)) at low temperature. 
We `calibrate' the $U/t$ of our model in terms of $r_R$
using $\Delta (r_R) \sim \Delta (U/t)$ in the insulating regime
(finite optical gap). We try a linear fitting
of this data and extrapolate it to lower $U/t$ values,
to have an estimate of the $U/t$ in terms of $r_R$
in the metallic regime (zero optical gap).
Our calibration (see figure \ref{gap}.(c)) suggests that 
for the rare-earth molybdates $U/t$ seems to vary from $\sim 5-9$ 
as R varies from Nd to Ho.

We now discuss results for the chosen $t$, $J_{AF}$,
and $U/t$, using absolute scales,  and compare with 
available experimental data
\cite{Mo-optics2,hanasaki_mag_expt}.

\begin{figure}[t]
\centerline{
\includegraphics[width=8.5cm,height=4.50cm]{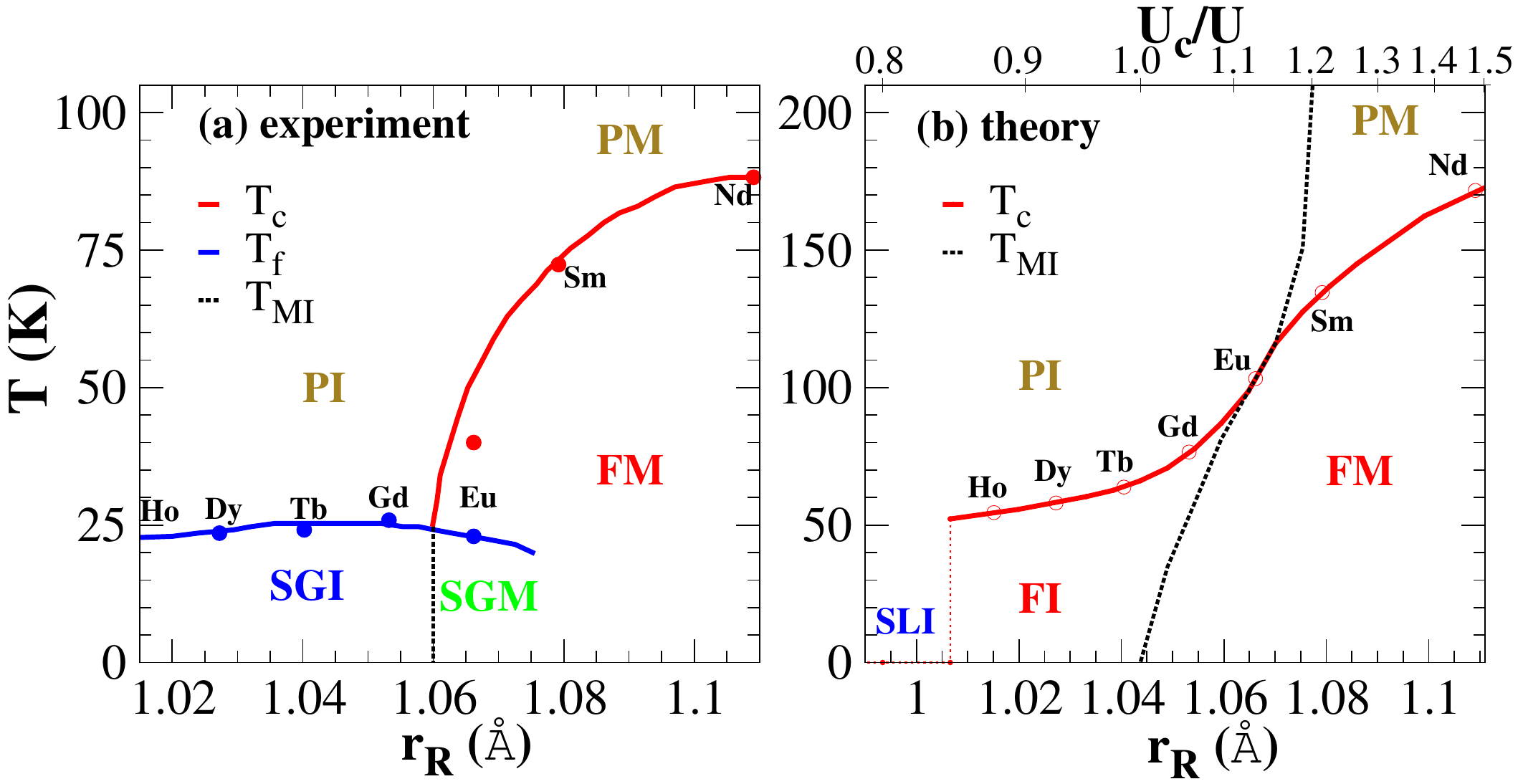}
}
\centerline{
\includegraphics[width=8.50cm,height=4.30cm]{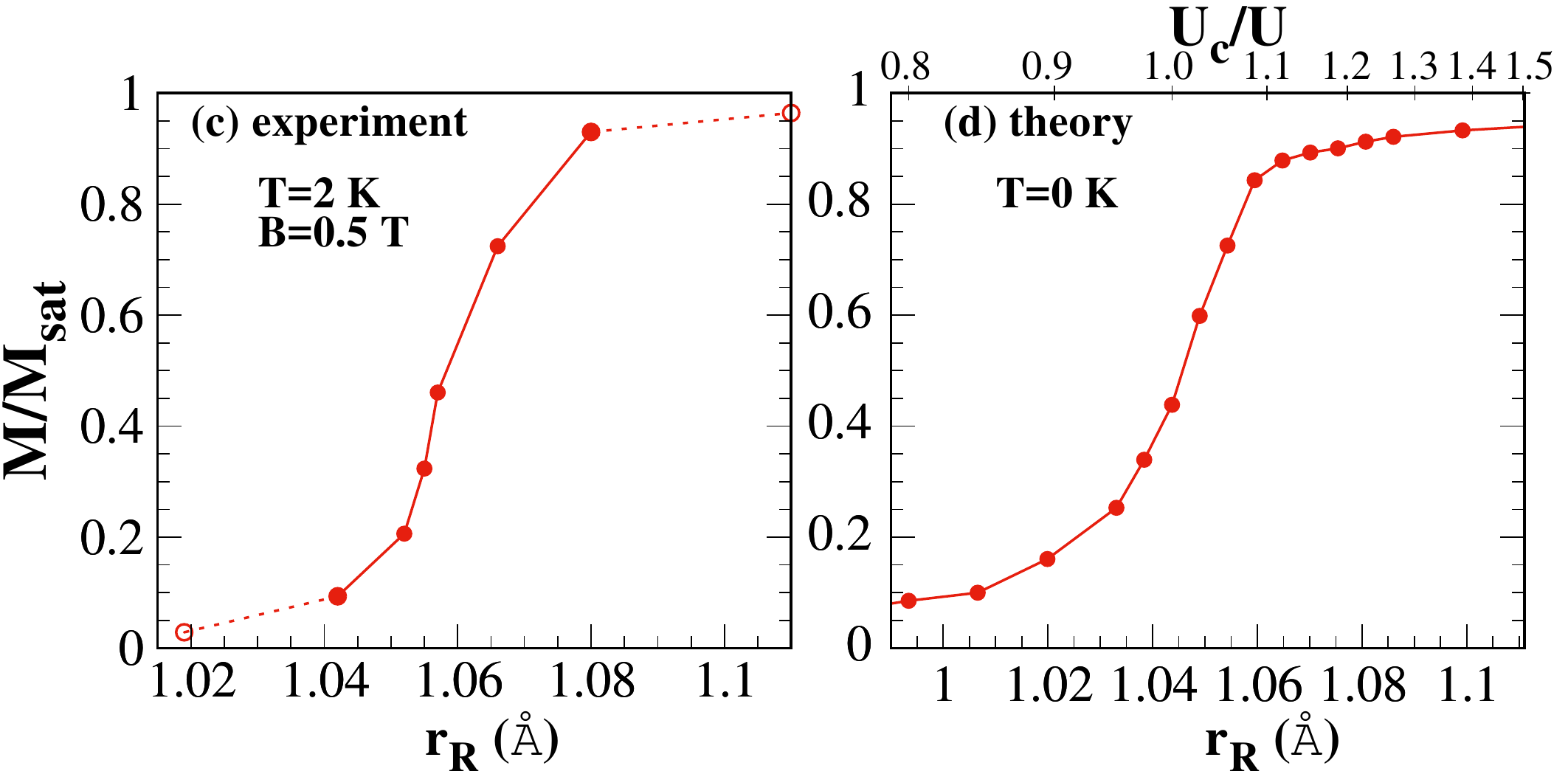}
}
\caption{\label{pd}Colour online: (a)-(b)~Phase diagram of the molybadates $-$
experiment \cite{Mo-anm-hall1,Mo-MIT-expt2} and theory.
The experimental ground state changes from
ferromagnetic metal (FM) to spin-glass metal (SGM) 
and then  spin-glass insulator (SGI) with reducing $r_R$.
Within theory the FM transforms to a `spin liquid' rather than 
a spin-glass. (c)-(d)~Show the ferromagnetic moment at low $T$ as
the system is taken through the MIT. 
Within both experiment \cite{hanasaki_mag_expt}, (c),
and theory, (d), a small moment survives in the insulator. 
In (b) we have cut off $T_c$ at the point where the $T=0$ 
magnetisation drops below $10 \%$.
}
\end{figure}

{\it Phase diagram:}
Fig.\ref{pd}.(a) shows the experimental phase diagram. 
At large $r_R$, where the $U/t$ ratio is relatively small, 
the ground state is a ferromagnetic metal 
with a moment $\sim 1.4 \mu_B$ per Mo 
\cite{Mo-anm-hall2,hanasaki_mag_expt}. 
The magnetisation seems to diminish slowly 
as $r_R$ reduces (panel \ref{pd}.(c)), and 
then rapidly around the metal-insulator transition, 
$r_R^c \sim 1.06 \AA $, but a small value 
survives into the weak insulating regime \cite{hanasaki_mag_expt}.
The FM $T_c$ is $\sim 80$K for large $r_R$ and drops sharply near $r_R^c$.
The state for $r_R \lesssim r_R^c$ is a spin glass, with
$T_{SG} \sim 20$K.

Panel \ref{pd}.(b) shows our result over the window $U/t \sim 5-10$. 
For our parameter calibration the 
metal-insulator transition (MIT) at $T=0$ 
occurs at $U_c \sim 7.6t$,
and we present our results in terms of $U_c/U$. 
At the right end, where $U \sim 0.7U_c$, 
the ground state is metallic, 
double exchange (DE) dominated,
and an almost  saturated ferromagnet. 
This is also a weakly `ferro orbital' state.
With increasing $U$ the orbital moment grows and leads to a
splitting of the $e'_g$ band. 

What drives the metal-insulator transition?
(i)~Increasing $U$ increases the splitting
$\Delta \sim \vert \Gamma \vert U$ between 
the electronic levels on Mo. This becoming comparable to the
bandwidth would lead to a Mott transition (the correlation aspect).
(ii)~With increasing $U$, 
the growing orbital moment suppresses the electron kinetic energy. 
This weakens DE. 
The competing AF superexchange reduces the magnetisation
and increases the extent of spin disorder in the ground state. 
We call this the `Anderson' (disorder) aspect of the problem. 
It depends crucially on the presence of $J_{AF}$.
$U_c$ is determined by a combination of 
the Mott and Anderson effects 
opening a gap in the electronic spectrum. 
Since magnetic disorder plays a role in the MIT, 
one can affect the transition by applying a magnetic field
\cite{hanasaki-andmott}.
Note that interaction effects are crucial in our problem 
in driving a metal-insulator transition. So we cannot
have a purely non interacting `Anderson transition'
without extrinsic disorder.

Panels \ref{pd}.(c) and \ref{pd}.(d) show
the low $T$ magnetisation in the molybdates and in
our scheme. The dependence is very similar and
a small magnetization survives beyond the MIT.

{\it Resistivity:}
We demarcate the finite $T$ metal-insulator boundary 
based on the temperature derivative of resistivity $d\rho/dT$:
`metal' if $d\rho/dT > 0$, `insulator' if $d\rho/dT < 0$. 
We compute $\rho(T)$ via the Kubo formula for changing $U/t$ 
\cite{kubo_allen}.

\begin{figure}[t]
\centerline{
\includegraphics[width=8.45cm,height=5.40cm]{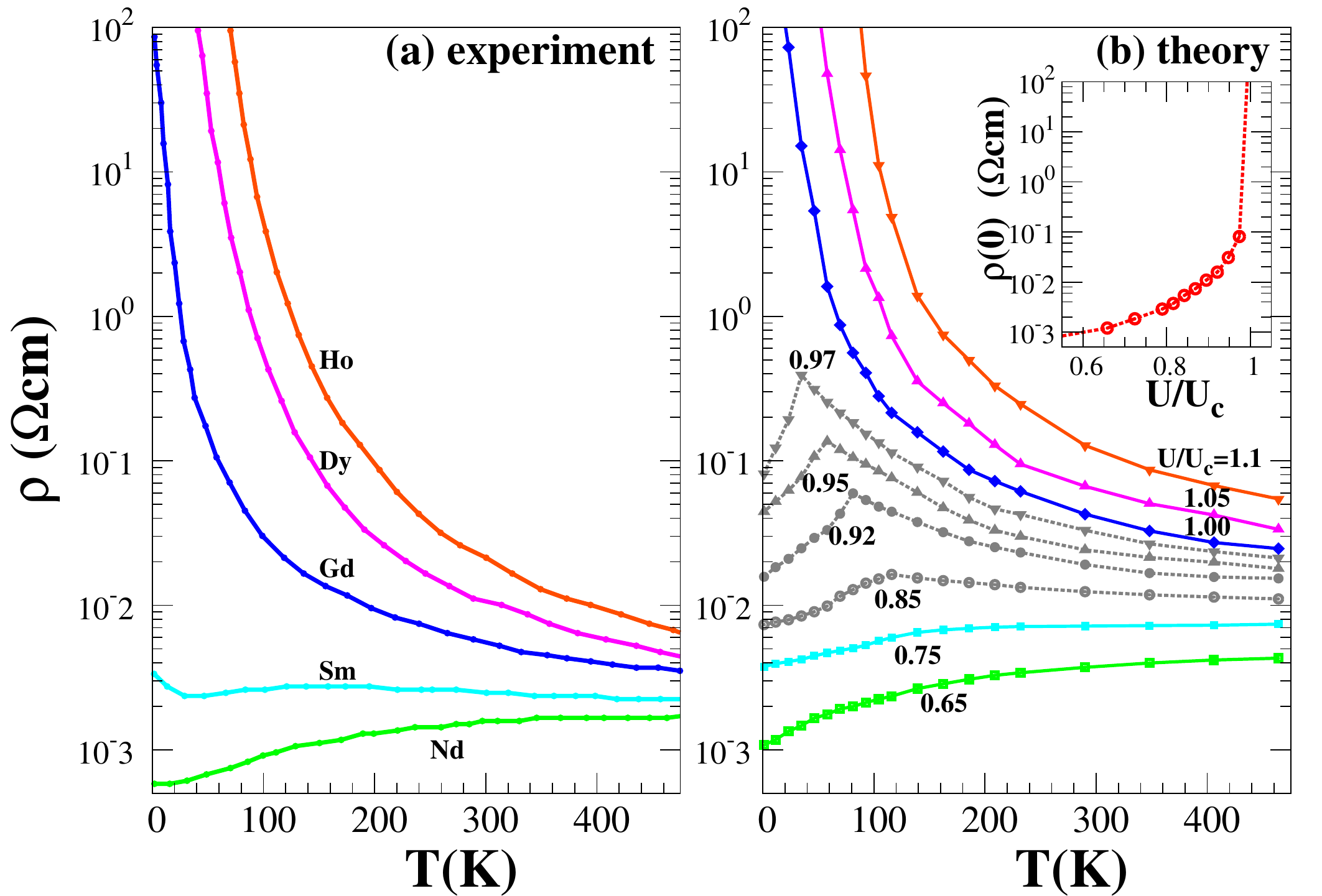}
}
\centerline{
\includegraphics[width=8.45cm,height=3.90cm]{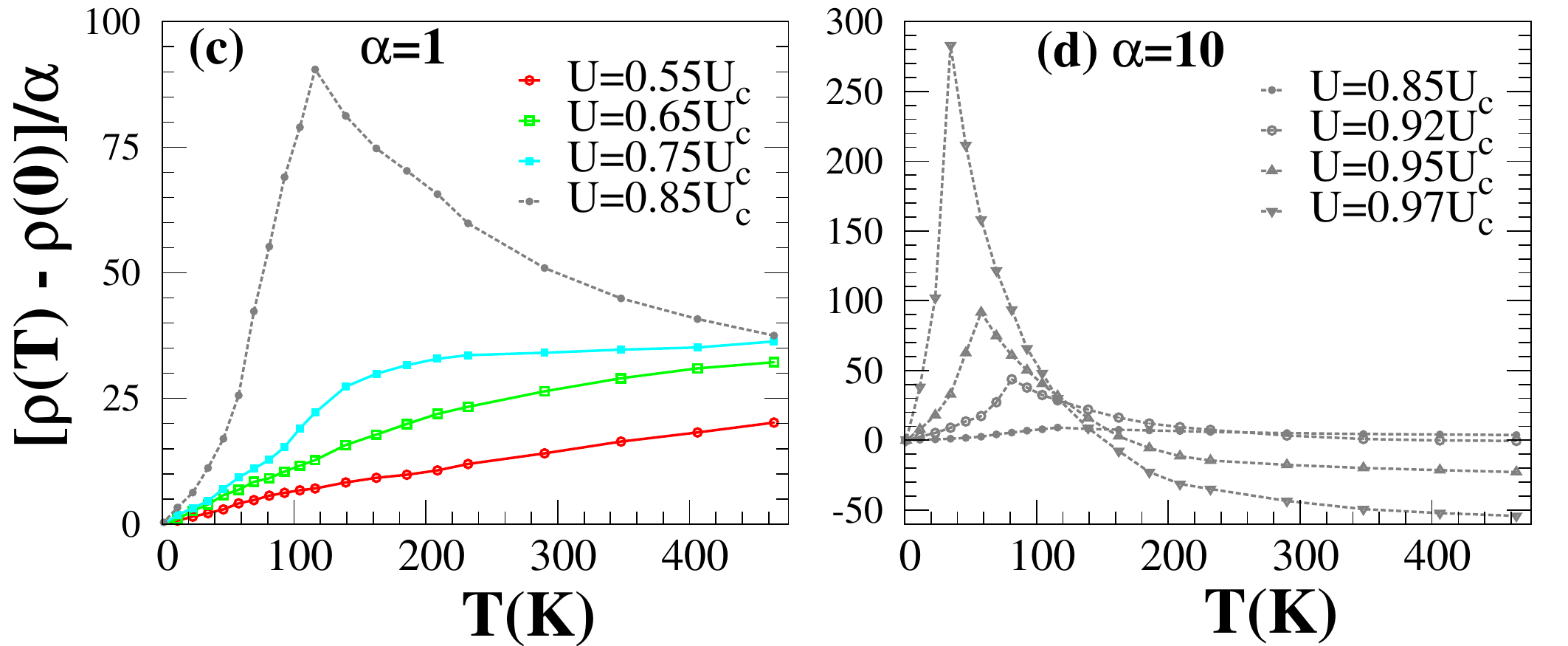}
}
\caption{\label{res}Colour online:
(a)~Resistivity of the molybdates for varying rare earth \cite{Mo-optics2}.
(b)~Resistivity computed within our scheme. The coloured plots
are counterparts of the experimental data, while the dotted lines
predict possible behaviour if intermediate compositions are to be
synthesized. Inset to (b) shows the growth of the `residual'
resistivity, $\rho(0)$. 
(c)-(d)~Show theory results for $\rho(T) - \rho(0)$ on a linear scale, 
to highlight the temperature driven metal to insulator crossover.}
\end{figure}

Fig.\ref{res}.(a) shows experimental resistivity \cite{Mo-optics2}
while \ref{res}.(b) shows the theory
result for parameter values set by the calibration.
Even the limited R variation in the experiments
can be thought to represent
three `regimes'. (i)~For R=Nd, the `high $T_c$' FM,
$\rho(T)$ has traditional metallic behaviour, 
$\rho(0) < 1$~m$\Omega$cm and $d\rho/dT > 0$ 
all the way to 400K. 
(ii)~For R=Gd,~Dy,~Ho, the system is  insulating
at all $T$, with $\rho(0) \rightarrow \infty$. The
behaviour is clearly activated for Dy and Ho while Gd seems
to be weakly insulating.
(iii)~R=Sm (and Eu, not shown) 
represents the most interesting case, with
$\rho(0) \sim 3$~m$\Omega$cm and a non monotonic $T$ dependence
\cite{Mo-optics2,Mo-optics4}.
Any theory would have to capture the obvious regimes (i) and (ii)
and also the peculiar large $\rho(0)$ and non monotonicity in (iii).

Our results, panel \ref{res}.(b), show the following:
$(i)$ For $U \ll U_c$, the itinerant $e'_g$ electrons
see a DE dominated ferromagnetic background,
as well as an orbital-ferro state. The $T=0$ state
is ideally clean, and finite $T$ resistivity 
from spin and orbital fluctuations generate an
approximate linear $T$ behaviour (see \ref{res}.(c)).
$(ii)$ For $U \gg U_c$, there is a distinct gap $\Delta$
with $\rho(T) \sim \rho_{0}e^{\Delta/T}$ as $T \rightarrow 0$
and $d\rho/dT < 0$ over the entire temperature range.
$(iii)$ For $U \lesssim U_c$, the residual resistivity $\rho(0)$
is finite. This arises from a combination of depleting DOS
at the Fermi level (due to the increasing orbital moment),
and the magnetic disorder due to weakening DE. The
behaviour of $\rho(0)$ is shown in the inset to panel \ref{res}.(b). 
Increasing $T$ does lead to a linear behaviour, with a large
slope, but the resistivity peaks at a scale $T_{peak}(U)$
and falls thereafter. Panels \ref{res}.(c)-(d) highlight this trend.
As $U \rightarrow U_c$, $T_{peak} \rightarrow 0$, 
finally merging with the insulating behaviour in $(ii)$.

\begin{figure}[t]
\centerline{
\includegraphics[width=4.3cm,height=3.7cm]{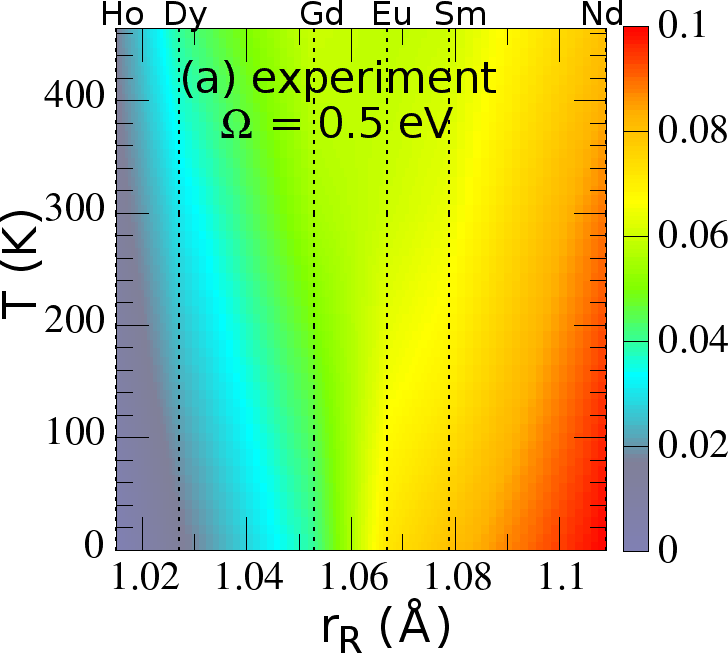}
\includegraphics[width=3.7cm,height=3.9cm]{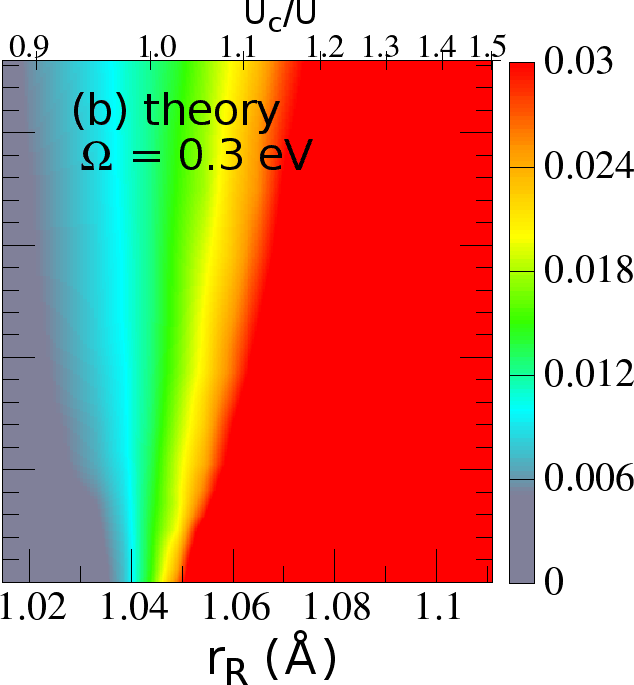}
}
\vspace{.2cm}
\centerline{
\includegraphics[width=4.3cm,height=3.7cm]{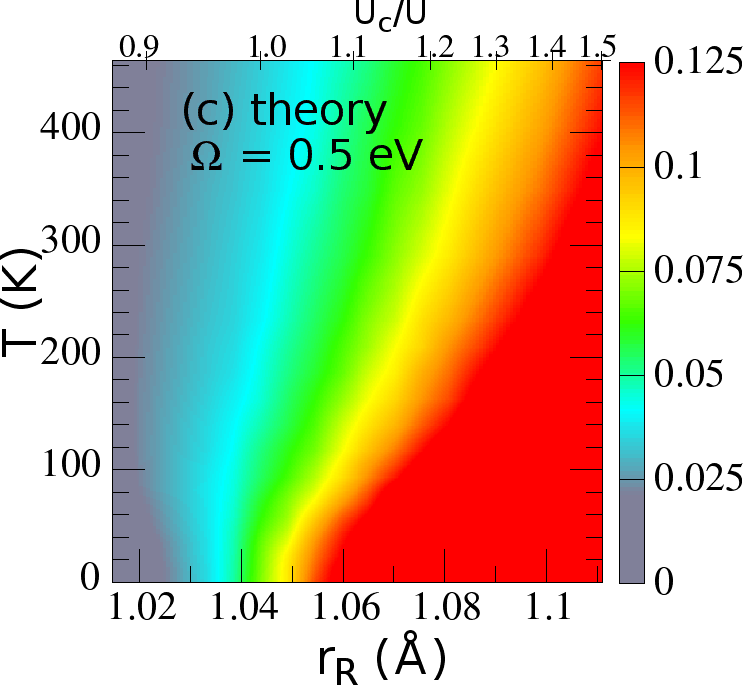}
\includegraphics[width=3.7cm,height=3.7cm]{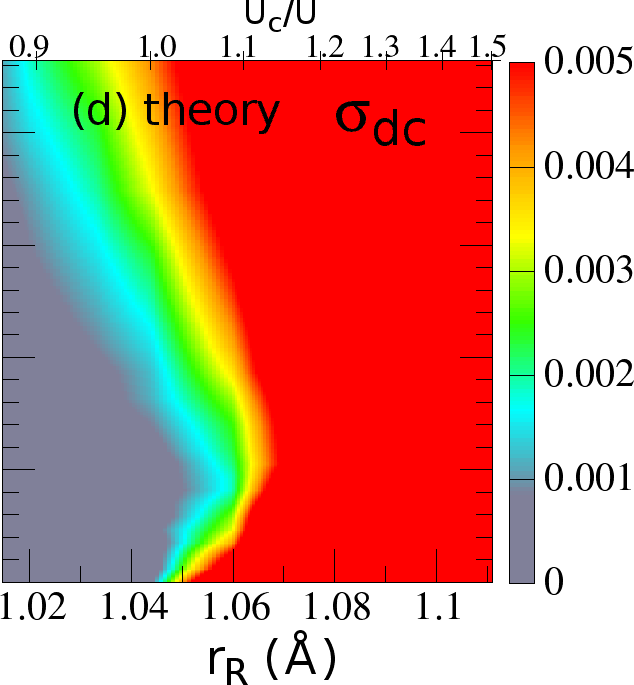}
}
\caption{\label{oc}Colour online:
(a)-(c) Low energy optical spectral weight,
$n_{eff} \propto \int_0^{\Omega} \sigma(\omega)d\omega$. 
(a)~Experimental result for $\Omega=0.5$~eV \cite{Mo-optics2}, 
(b)~theory result for $\Omega=0.3$~eV,
(c)~theory result for $\Omega=0.5$~eV. 
(d)~The d.c conductivity within theory.
}
\end{figure}

{\it Optical spectral weight:}
Fig.\ref{oc}.(a) 
shows the experimentally estimated optical spectral weight
$n_{eff}(\Omega) = ({2m_0}/{\pi e^2}) 
\int_0^{\Omega} \sigma(\omega) d\omega$ for different R and
varying temperature at $\Omega=0.5$~eV \cite{Mo-optics2}.
It shows the expected trend of 
$n_{eff}$ growing with $T$ in the insulating, low $r_R$, side
as the Mott gap is slowly filled, and reducing on the metallic
side as weight gets transferred to high energy as coherence is
lost. 

We calculated the same quantity for different
cutoff frequencies, $\Omega$, as
$n_{eff}(\Omega) = \int_{0}^{\Omega} \sigma(\omega) d\omega$.
Fig.\ref{oc}.(b)-(c) show our result for $\Omega=0.3$~eV 
and $0.5$~eV respectively. 
Panel \ref{oc}.(d) shows just $\sigma_{dc}$ 
to contrast the features in optical weight to
the non monotonicity of the d.c conductivity itself.

Our result at $\Omega=0.3$~eV, roughly 2/3 of the experimental
cutoff, has the same features as the experimental data. At 
$\Omega = 0.5$~eV, however, our data reveal a weak non 
monotonicity in the $T$ dependence when $ U \gtrsim U_c$.
This arises because $\sigma(\omega)$ gains weight at
low frequency, as in panel (b), but {\it loses more}
around $\omega \sim 0.5$~eV.  
The success in capturing the d.c resistivity, Fig.\ref{res}, 
does not translate to a similar success 
in capturing the high energy optical conductivity.
It is possible that some of the simplifying assumptions 
regarding bandstructure and coupling constants, {\it i.e}
$J_H$, affect this result. 

The non monotonicity in our $\Omega =0.5$ eV spectral weight
(Fig.\ref{oc}.(c)) is distinct from the d.c conductivity behaviour shown
in Fig.\ref{oc}.(d). Fig.\ref{res}.(b) shows that bad $T=0$ metals, 
for $U \lesssim U_c$, 
become more resistive with increasing $T$ and beyond a
$T_{peak}$ become less resistive again. We suggest that
a detailed conductivity map, on materials like
Gd$_{2-x}$Sm$_x$Mo$_2$O$_7$ or Gd$_{2-x}$Eu$_x$Mo$_2$O$_7$
could reveal this non monotonicity.

\begin{figure}[t]
\centerline{
\includegraphics[width=5.3cm,height=3.3cm]{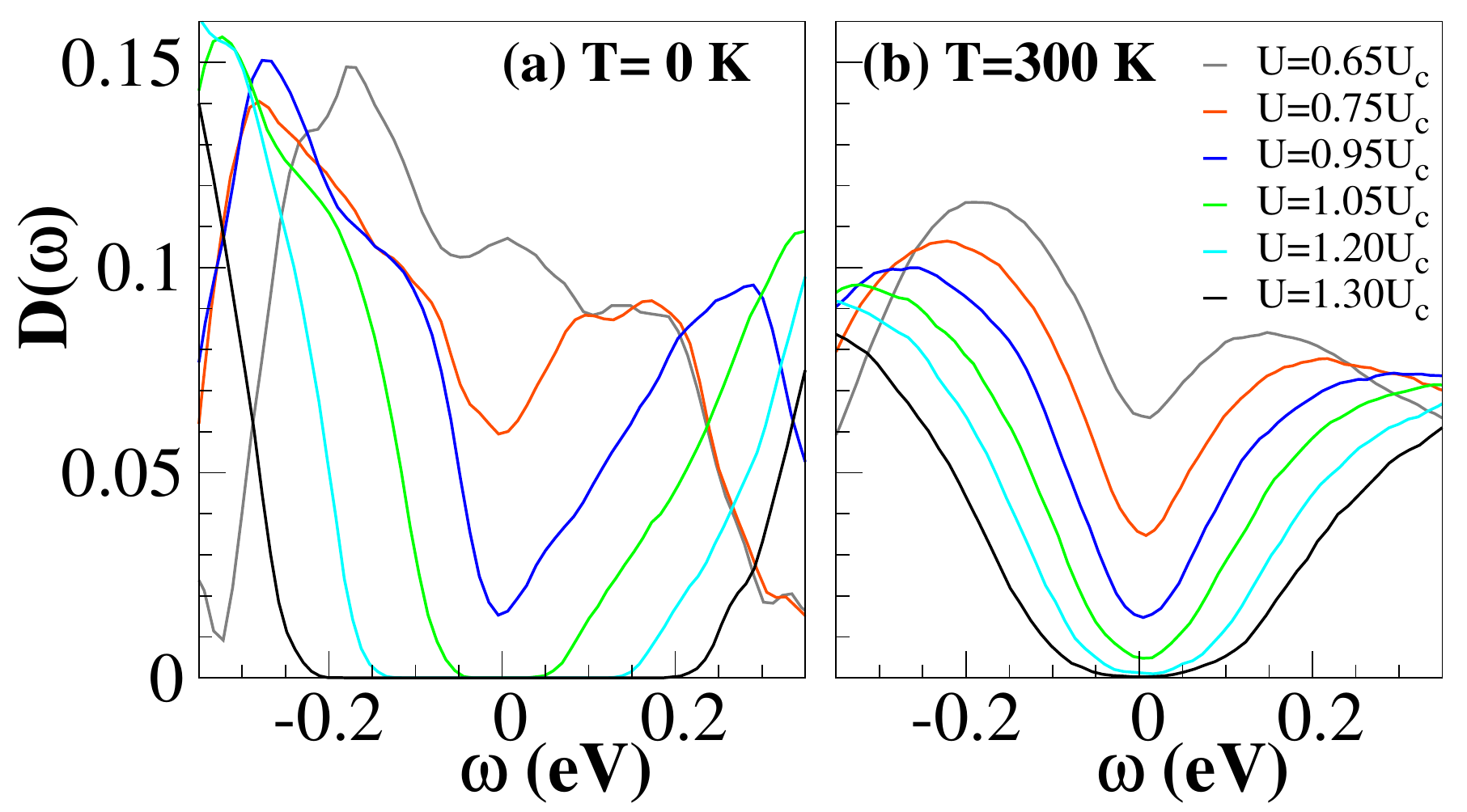}
\includegraphics[width=3.2cm,height=3.3cm]{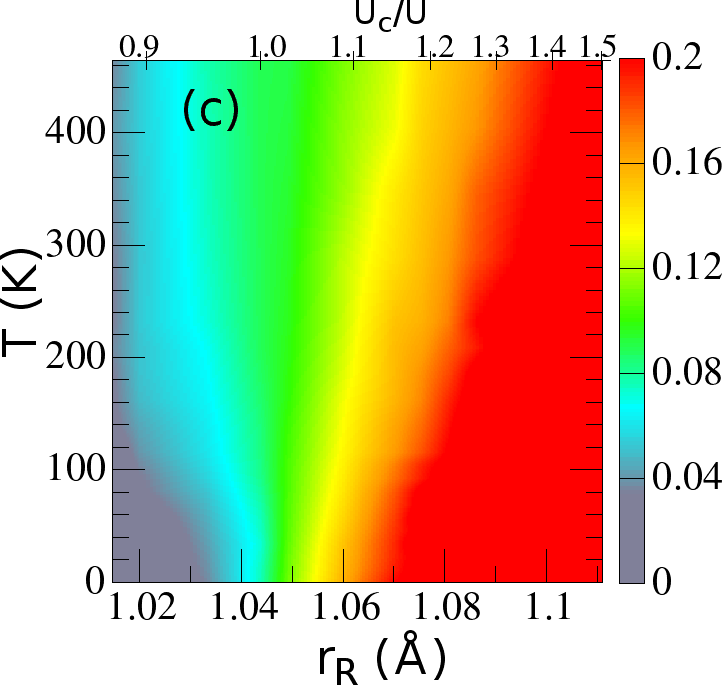}
}
\centerline{
\includegraphics[width=8.5cm,height=3.3cm]{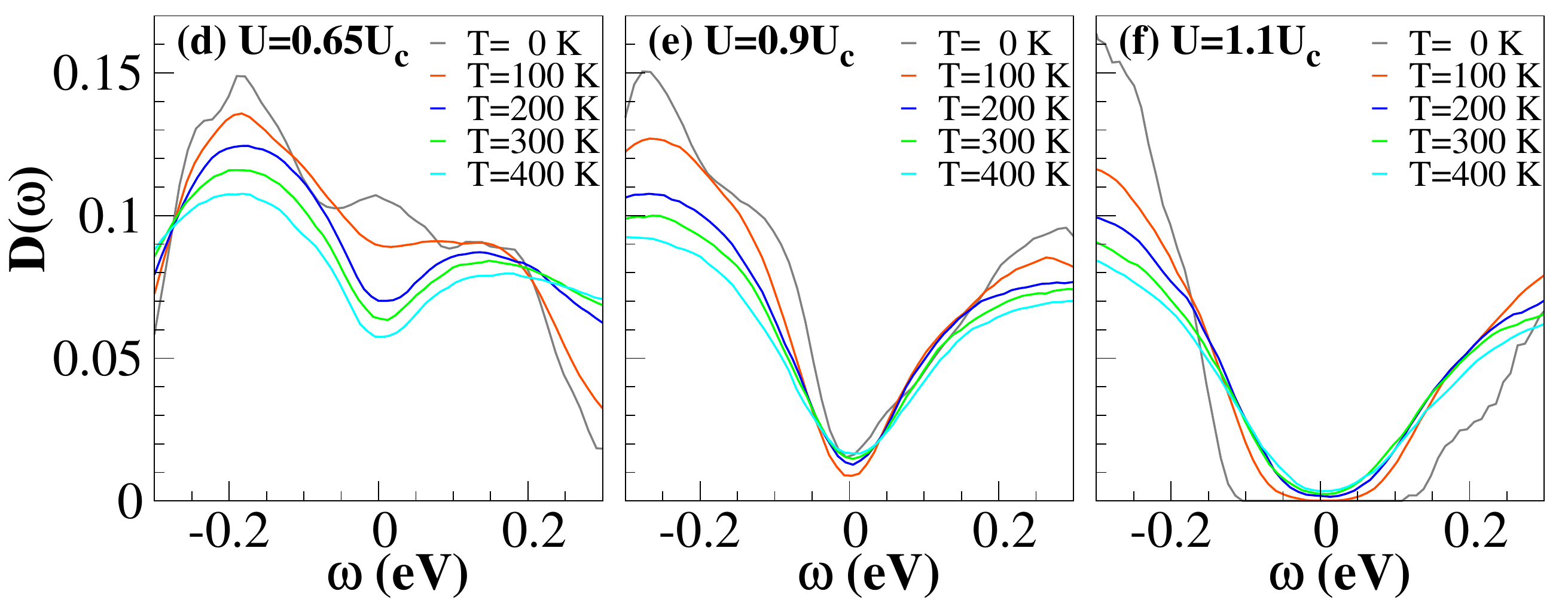}
}
\caption{\label{dos}Colour online:
(a)-(b)~Variation of density of states with $U$
at $T=0$K and $300$K. 
(c)~Integrated DOS, over $\omega =\pm 0.15$eV, 
for varying $U$ and $T$.
(d)-(f)~Temperature dependence of DOS at $U=0.65U_c$,
$0.9U_c$ and $1.1U_c$, respectively.
Panel (d) shows a gapless metal,
(e)~shows a pseudogapped state, 
while (f)~shows the $T$-dependence in a gapped Mott insulator.
}
\end{figure}

{\it Density of states:}
We computed the single particle density of states (DOS),
$D(\omega) = {1 \over N} \sum_n \langle \delta(\omega - \epsilon_n) \rangle$, 
for the interaction and temperature window studied.
Fig.\ref{dos}.(a) shows the dependence of $D(\omega)$ on $U/U_c$
as the system is driven across the Mott transition at $T=0$.
The DOS has its tight binding form upto
$U \sim 0.7U_c$ beyond which the presence of the orbital
moment shows a visible depletion in the DOS around $\omega =0$.
This dip becomes a gap for $U \ge U_c$, 
which grows in the insulating phase.
At $T=300$K, Fig.\ref{dos}.(b), the systems with $U < U_c$ {\it lose
weight} near $\omega=0$, while those with $U > U_c$ gain
weight. Panel \ref{dos}.(c) quantifies these trends by 
calculating $\int_{- \Omega'}^{\Omega'} D(\omega)d \omega$,
where $\Omega' =0.15$~eV (to make a comparison with Fig.\ref{oc}.(b)).
We suggest that the optical behaviour observed experimentally
has an analog in the single particle spectral weight transfer as
well.

Figures \ref{dos}.(d)-(f) show the thermal evolution of the DOS
at three representative $U/U_c$.
$(i)$ In fig. \ref{dos}.(d), for $U \sim 0.6U_c$, 
the ground state is a nearly saturated ferromagnet 
with a small orbital moments ${\bf \Gamma}_i$'s and
has finite DOS at $\omega=0$.
Thermal growth and fluctuations of the ${\bf \Gamma}_i$'s 
decrease the DOS at $\omega=0$ resulting in 
a small dip at high temperature. 
$(ii)$ For $U=1.1U_c$, fig. \ref{dos}.(f), 
there is significant spin disorder in the ground state
(the ferromagnetic moment is $\sim 0.1$) and 
the ${\bf \Gamma}_i$'s are large, $\sim 1$, at all sites. 
A remnant of the atomic gap $\sim U \vert \Gamma \vert$
survives despite the presence of hopping.
The DOS shows a Mott gap.
With increase in temperature, the angular fluctuations 
of the ${\bf \Gamma}_i$'s result in a slight smearing of the
gap edge and increase in `low energy' weight.
$(iii)$ For $U=0.9U_c$, fig. \ref{dos}.(e), 
the magnetic state has magnetisation, $M \sim 0.5$ 
and the ${\bf \Gamma}_i$'s are moderately large.
As a result there is only a loss in weight around $\omega =0$
but no hard gap. This is a pseudogap state. 

\section{Discussion}

There are some issues where our modeling differs from the
experimental results. The origin of these differences is
obvious, and we briefly touch upon them:
(i)~{\it Temperature scales:}
We have used a model with Hund's coupling 
$J_H \gg t $ for convenience, 
and obtain $T_c \sim 160$K for Nd. 
In reality $J_H \sim 5t$ \cite{molybdate-DFT1},
and as earlier results show \cite{kalpataru_DE}
this would reduce $T_c$ by $\sim 50 \%$ 
to about $80$K, close to the actual value for Nd.
(ii)~{\it Spin freezing:}
Within our scheme we do not find any spin freezing, 
so no $T_{SG}$. Our magnetic state for $U \gtrsim U_c$ 
is a spin liquid (SL), rather than a spin glass,
with weak ferromagnetism. To obtain freezing, and the
correct $T_{SG}$, it seems that one requires significant
bond disorder \cite{pyr_HAF_SG} (or bond distortions
\cite{Molyb-els-coup-NMR,Molyb-els-coup-muon,Molyb-els-coup-neutron,
Molyb-els-coup-th}).
(iii)~We have also not touched upon
the experimental $P-r_R$ phase diagram,
the field driven 
IMT in Gd$_2$Mo$_2$O$_7$ \cite{hanasaki-andmott}, 
and the anomalous Hall effect (AHE)
observed in Nd$_2$Mo$_2$O$_7$ 
\cite{Mo-anm-hall1,Mo-anm-hall2,Mo-anm-hall3, Mo-anm-hall4,Mo-anm-hall5}. 

\section{Conclusion}
We provide the first study of the Mott transition in the pyrochlore
molybdates, in a real space framework, retaining the double 
exchange, superexchange and correlation effects.
Our phase diagram captures the transition from a 
ferromagnetic metal to a spin disordered insulator, as in
experiments, with reasonable thermal scales. Our 
transport and optical results capture major features
of the molybdate data, reveal unexplored non
monotonicities, and, we predict, correlate with 
the single particle spectral weight.
Ongoing work will address the field driven Anderson-Mott transition.

\acknowledgments
We acknowledge use of the HPC clusters at HRI.


\begin{thebibliography}{0}
  
\bibitem{MIT_Mott} 
  MOTT N. F., Proc. Roy. Soc. A {\bf 62}, (1949) 416;\\
  MOTT N. F., {\it Metal-Insulator Transitions}, (London: Taylor and Francis) (1990).  
  
\bibitem{SL-triangle}%
  ANDERSON P. W., 
  Mater. Res. Bull. {\bf 8}, (1973) 153.
  
\bibitem{SL-fr-mag}
  BALENTS L., 
  Nature (London) {\bf 464}, (2010) 199.
  
\bibitem{SL-tr-organic1}
  SHIMIZU Y., MIYAGAWA K., KANODA K., MAESATO M., and SAITO G.,  
  Phys. Rev. Lett. {\bf 91}, (2003) 107001.
  
\bibitem{SL-tr-organic2}
  YAMASHITA M., et al. 
  Nature Phys. {\bf 5}, (2009) 44.
  
\bibitem{SL-tr-organic}  
  KANODA K. and KATO R., 
  Annu. Rev. Condens. Matter Phys. {\bf 2}, (2011) 167.
  
\bibitem{pyr-rmp}  
  GARDNER J. S., GINGRAS M. J. P., and GREEDAN J. E., 
  Rev. Mod. Phys. {\bf 82}, (2010) 53.
    
\bibitem{Mo-var-mag-ele1}
  GREEDAN J. E., {\it et al.}, 
  J. Solid State Chem. {\bf 68}, (1987) 300.
  
\bibitem{Mo-var-mag-ele2}
  MIYOSHI K., {\it et al.}, 
  J. Magn. Magn. Mater. {\bf 226}, (2001) 898.

\bibitem{Mo-optics2}%
  KEZSMARKI I., {\it et al.},
  Phys. Rev. Lett. {\bf 93}, (2004) 266401.
  
\bibitem{Mo-MIT-expt2}
  IGUCHI S., {\it et al.},
  Phys. Rev. Lett. {\bf 102}, (2009) 136407.  
  
\bibitem{Mo-optics4}%
  KEZSMARKI I., {\it et al.},
  Phys. Rev. B {\bf 73}, (2006) 125122.  
  
  \bibitem{Mo-Tc-Tf1}
  ALI N., {\it et al.}, 
  J. Solid State Chem. {\bf 83}, (1989) 178.
\bibitem{Mo-Tc-Tf2}
  ALI N., {\it et al.},
  J. Alloys Compd. {\bf 181}, (1992) 281.
\bibitem{Mo-Tc-Tf3}
  TAGUCHI Y. and TOKURA Y., 
  Phys. Rev. B {\bf 60}, (1999) 10280.  
  
\bibitem{Mo-anm-hall1}%
  KATSUFUJI T., HWANG H.Y., and CHEONG S.W., 
  Phys. Rev. Lett. {\bf 84}, (2000) 1998.
\bibitem{Mo-anm-hall2}%
  TAGUCHI Y., {\it et al.},
  Science {\bf 291}, (2001) 2573.
\bibitem{Mo-anm-hall3}%
  TAGUCHI Y., {\it et al.},
  Phys. Rev. Lett. {\bf 90}, (2003) 257202.
\bibitem{Mo-anm-hall4}%
  IGUCHI S., HANASAKI N., and TOKURA Y.,
  Phys. Rev. Lett. {\bf 99}, (2007) 077202.
\bibitem{Mo-anm-hall5}%
  UEDA K., {\it et al.},
  Phys. Rev. Lett. {\bf 108}, (2012) 156601.
  
\bibitem{hanasaki-andmott}
  HANASAKI N., {\it et al.},
  Phys. Rev. Lett. {\bf 96}, (2006) 116403. 
  
\bibitem{molybdate-DFT1}
  SOLOVYEV I. V., 
  Phys. Rev. B {\bf 67}, (2003) 174406.

\bibitem{molybdate-DFT2}
  SHINAOKA H., MOTOME Y., MIYAKE T., and ISHIBASHI S.,
  Phys. Rev. B {\bf 88}, (2013) 174422.
  
\bibitem{Molyb-init-model}%
  MOTOME Y. and FURUKAWA N.,
  J. Phys.: Conf. Ser. {\bf 320}, (2011) 12060. 
  
\bibitem{Molyb-DE-SE}%
  MOTOME Y. and FURUKAWA N.,
  Phys. Rev. Lett. {\bf 104}, (2010) 106407,
  Phys. Rev. B {\bf 82}, (2010) 060407(R).    
  
\bibitem{DMFT-georges}  
  GEORGES A., {\it et al.}, 
  Rev. Mod. Phys. {\bf 68}, (1996) 13.
  
\bibitem{hubb-strat}
  HUBBARD J., 
  Phys. Rev. Lett. {\bf 3}, (1959) 77.

\bibitem{hubbard}
  HUBBARD J., 
  Phys. Rev. B {\bf 19}, (1979) 2626.

\bibitem{schulz} 
  SCHULZ H. J., 
  Phys. Rev. Lett. {\bf 65}, (1990) 2462.

\bibitem{method_SPA_detail} 
 SWAIN N. and MAJUMDAR P., 
 arXiv:1610.00695 (2016).

\bibitem{SPA_dagotto}%
  MAYR M., ALVAREZ G., SEN C., and DAGOTTO E.,
  Phys. Rev. Lett. {\bf 94}, (2005) 217001.

\bibitem{SPA_meir}  
  DUBI Y., {\it et al.}, 
  Nature, {\bf 449}, (2007) 876.

\bibitem{mott-Tr} 
  TIWARI R. and MAJUMDAR P., 
  Europhys. Lett. {\bf 108}, (2014) 27007.

\bibitem{tca-sanjeev-pinaki} 
  KUMAR S. and MAJUMDAR P., 
  Eur. Phys. J. B, {\bf 50}, (2006) 571.
  
\bibitem{tca-anamitra-dagotto} 
  MUKHERJEE A., PATEL N.D., BISHOP C., and DAGOTTO E., 
  Phys. Rev. E {\bf 91}, (2015) 063303.

\bibitem{kubo_allen}
  ALLEN P. B. in {\it Conceptual Foundation of Materials V.2},
  edited by LOUIE S. G. and COHEN M. L., Elsevier (2006).

\bibitem{hanasaki_mag_expt}
  HANASAKI N., {\it et al.},
  Phys. Rev. Lett. {\bf 99}, (2007) 086401.
  
\bibitem{kalpataru_DE} %
  PRADHAN K. and MAJUMDAR P., 
  Europhys. Lett. {\bf 85}, (2009) 37007.

\bibitem{pyr_HAF_SG} 
  SAUNDERS T. E. and CHALKER J. T.,
  Phys. Rev. Lett. {\bf 98}, (2007) 157201.

\bibitem{Molyb-els-coup-NMR}%
  KEREN A. and GARDNER J. S., 
  Phys. Rev. Lett. {\bf 87}, (2001) 177201.

\bibitem{Molyb-els-coup-muon}%
  SAGI E., OFER O., KEREN A., and GARDNER J.S.,
  Phys. Rev. Lett. {\bf 94}, (2005) 237202. 
  
\bibitem{Molyb-els-coup-neutron}%
  GREEDAN J. E., {\it et al.},
  Phys. Rev. B {\bf 79}, (2009) 014427. 

\bibitem{Molyb-els-coup-th}%
  SHINAOKA H., TOMITA Y., and MOTOME Y., 
  Phys. Rev. Lett. {\bf 107}, (2011) 047204, 
  Phys. Rev. B {\bf 90}, (2014) 165119.


\end{thebibliography}
\end{document}